\documentclass[journal=jpcl,manuscript=article]{achemso}

\usepackage{amsmath}
\usepackage{graphicx}
\usepackage{epsfig}
\usepackage{color}

\def\beq{\begin{equation}}
\def\eeq{\end{equation}}
\def\beqa{\begin{eqnarray}}
\def\eeqa{\end{eqnarray}}

\def\dag{\dagger}

\newcommand{\appropto}{\mathrel{\vcenter{
			\offinterlineskip\halign{\hfil$##$\cr
				\propto\cr\noalign{\kern2pt}\sim\cr\noalign{\kern-2pt}}}}}

\author{Elinor Zerah-Harush}
\author{Yonatan Dubi}
\email{jdubi@bgu.ac.il}
\affiliation{Department of Chemistry and Ilse-Katz Institute for Nanoscale Science and Technology, Ben-Gurion University of the Negev, Beer-Sheva 84105, Israel}

	\title{Universal Origin for Environment-Assisted  Quantum Transport in Exciton Transfer Networks} 
\begin{document}  % HERE FOR ACS 
	\begin{abstract}
    Environment-assisted quantum transport (ENAQT) is the possibility of an external environment to enhance transport efficiency of quantum particles. This idea has generated much excitement over recent years, especially due to the experimentally-motivated possibility of ENAQT in photo-synthetic exciton transfer complexes. Many theoretical calculations have shown ENAQT, but the explanations for its origin differ, and a universal explanation has been elusive. Here we demonstrate a universal origin for ENAQT in quantum networks with a dephasing environment, based on a relation between exciton current and occupation within a Markovian open quantum system approach. We show that ENAQT appears due to two competing processes, namely the tendency of dephasing to make the exciton population uniform, and the formation of an exciton density gradient, defined by the source and the sink. Furthermore, we find a geometric condition on the network for the appearance of ENAQT, relevant to natural and artificial systems.    
    
    \end{abstract}
	\maketitle   % ONLY FOR PHYS REV

%\section{Introduction}
Photosynthesis is the process used by plants, algae and some bacteria to convert solar energy into chemical energy. Different species perform photosynthesis with different machinery, but despite the differences, a good model for many photosynthetic complexes is a general three-part structure comprising an antenna, a reaction center, and an exciton transfer complex (ETC), which connects the two. The ETC is a network of chromophores embedded in a protein environment, and its structure and size vary from one organism to another. Its function, however, is similar: transporting the excitation energy from the antenna to the reaction center.\cite{mohseni2014quantum}

It has been accepted for many years that excitons are transported by an incoherent  hopping process, i.e. by classical diffusive dynamics, via the $F\ddot{o}rster$ resonance energy transfer \cite{govorov2016}. While this seems the case for many molecular aggregates ~\cite{levi2015}, it is apparently not the whole picture in light-harvesting complexes (LHC). For instance, it cannot explain the high efficiency of those complexes \cite{fleming2004,levi2015}. Moreover, it cannot explain results of ultrafast nonlinear spectroscopy experiments \cite{engel2007,Calhoun2009,Collini2010,Panitchayangkoon2011}, showing evidence for long-lived oscillatory signals that were conjectured to be of quantum mechanical origin \cite{mohseni2014quantum}. 

In natural transfer complexes, the chromophore network that the exciton is transported through is covered by a large protein structure. Exposed to room temperature, it is far from the "clean" environment in which quantum systems are traditionally described. Consequently, the idea that a biological system, that apparently exploit all of its resources, may be using quantum coherence as a resource to assist energy transport has generated much excitement (see, e.g., reviews in  Refs.~\cite{Ishizaki2012,Collini2013,Lambert2013,Pachon2012} and many references therein). The central phenomenon behind this idea is "environment assisted quantum transport" (ENAQT). According to the principle of ENAQT, the environment interrupts the phase-coherent transport of the quantum-mechanical excitations through the transfer complex by dephasing, in a way that {\sl enhances} the efficiency of the energy transport.  While the role of quantum mechanical transport and ENAQT in photosynthetic systems is still under debate\cite{kassal2013,Miller2012,ritschel2011,duan2017nature}, the concept of ENAQT  is not limited to photosynthetic complexes, and was expanded to, e.g., electronic and molecular systems \cite{Semiao2010,Nalbach2010, Scholak2011a,Ajisaka2015,Lim2014,leon2015noise}, cold atoms \cite{Scholak2011a} and photonic crystals \cite{Biggerstaff2016,viciani2016disorder,Caruso2016}.

%\doublespacing
ENAQT may arise from different mechanisms, and various suggestions were made in the theoretical literature to explain it. %\cite{Rebentrost2009,Mohseni2008,Caruso2009,Chin2010,Plenio2008,Kassal2012,Caruso2014,Li2015,Marais2013,Chen2013,dubiinterplay2015,Nesterov2013,Berman2015,Baghbanzadeh2016,wu2010efficient,cao2009optimization,PhysRevLett.110.200402,Hoyer2010,sarovar2010}.
These include dephasing-induced delocalization \cite{Rebentrost2009, Manzano2013, Chin2010,cao2009optimization}, momentum rejuvenation \cite{Li2015}, opening and broadening transport channels \cite{Plenio2008, Caruso2009,wu2010efficient}, line-broadening \cite{Caruso2009}, superradiance \cite{Nesterov2013,Berman2015}, supertransfer and funneling \cite{Baghbanzadeh2016}, trapping-time crossover \cite{PhysRevLett.110.200402}, directed flow \cite{dubiinterplay2015} and more. Here, we suggest that there is a universal mechanism for the origin of ENAQT in quantum networks which are larger than two chromophores (the realistic situation) and have a broad vibrational spectrum (also a realistic situation), making our theory relevant to many of the models presented in the literature (and, presumably, to natural ETCs).  Specifically, we  show that exciton transfer enhancement is a result of two competing processes, namely the tendency to uniformly spread the exciton population along the network, and the formation of a uniform population gradient.  

%\section{Results}
Our results and analysis are based on calculation of exciton currents through a quantum network, defined by the general tight-binding Hamiltonian
\begin{equation}\label{eq:Tb_Of_Chain}
H=\sum_{i=1}^{n} \epsilon_i a_i^{\dagger}a_i-\sum_{i,j=1}^{n} t_{i,j} a_i^{\dagger}a_{j}~,
\end{equation}
where $a^\dagger$ and $a_i$ are creation and annihilation operators of an exciton at chromophore \textit{i}, $\epsilon_i$ are on-site exciton energies, and $t_{i,j}$ are coupling elements
between two chromophores. The (Markovian) dynamics are described by the Lindblad equation \cite{breuer2002theory}:
\begin{equation}\label{eq: 31}
\frac{d\rho_S}{dt}=-i[H_S,\rho_S]+L\rho_s ~~,
\end{equation}

where $\rho_S$ is the density matrix of the reduced system and $L$ is the Lindbladian, defined as 
\begin{equation}
L\rho_s=\sum_{k} \Gamma_k\big(V_k\rho_S V_k^\dag-\frac{1}{2}\{V_k^\dag V_k,\rho_S\}\big)~~,
\end{equation}
where $V_k$ are Lindblad operators describing the action of the environment on the system, and $\Gamma_k$ is the respective rate of the Lindblad operator. The index $k$ represents different environments and/or different processes induced by these environments on the system. 

Here, we consider the quantum network to be in contact with a source (the antenna), a sink (reaction center), and a dephasing channel (ETC protein environment), characterized by an exciton injection rate $\Gamma_{inj}$, extraction rate $\Gamma_{ext}$ and dephasing rate $\Gamma_{deph}$, respectively. In the presence of these environments, the Lindblad equation (Eq. ~\ref{eq: 31} ) has the form
\begin{equation}\label{eq:5001}
\frac{d\rho_s}{dt}=-i[H,\rho_s]+L_{inj}\rho_s+L_{ext}\rho_s+L_{dep}\rho_s ~.
\end{equation}
$L_{inj}$, $L_{ext}$, $L_{dep}$ are the injection, extraction and dephasing elements, respectively, with the corresponding operators $V_{inj}=a^{\dagger}_{inj},~V_{ext}=a_{ext}$ describing creation and anihilation of an exciton in the injection and extraction sites. These operators describe the non-equilibrium condition in which energy is constantly pumped into the system, but not the equilibrium limit \cite{gelbwaser2017thermodynamic}. Calculations were also performed with a thermodynamically consistent model \cite{gelbwaser2017thermodynamic,Dubi2009d}, and the results are similar (see SI for details). The dephasing operator is a local measurement, $V_{dep,i}= a^\dagger_i a_i$, and the dephasing part of the Lindbladian is $L_{dep}\rho_s=\sum_i L_{dep,i}\rho_s$, where $L_{dep,i}\rho_s=\Gamma_{deph}\big(V_{dep,i}\rho_S V_{dep,i}^\dag-\frac{1}{2}\{V_{dep,i}^\dag V_{dep,i},\rho_S\}\big)$. This form ensures that the fluctuations (and the ensuing dephasing) are local to each chromophore and not correlated between different chromophores. 

We proceed by calculating the exciton transport at the steady-state, which seems to be the relevant state for natural systems\cite{manzano2012,dubiinterplay2015}. Bothe energy and exciton currents can be evaluated by noting that the total energy, $\bar{E}=\mathrm{Tr} \left( H\rho_s \right)$ and total exciton number,$\bar{N_e}=\mathrm{Tr} \left( \hat{n} \rho_s \right)$  (where $ \hat{n}=\sum_i a^\dagger_ia_i$ is the total number operator), are time-independent, which allows for a propeper definiton of the energy and exciton currents, $J_q=\mathrm{Tr} \left( H L_{ext} \right),J_p=\mathrm{Tr} \left( \hat{n} L_{ext} \right)$   \cite{dubiinterplay2015, manzano2012}.  Solving for the steady state, i.e. $\frac{d\rho_s}{dt}=0$, one finds for the heat current
\begin{equation}\label{eq:16}
J_q=\Gamma_{ext}[\epsilon_n\rho_{n,n}+\frac{1}{2}t(\rho_{n,n-1}+\rho_{n-1,n})]~~,
\end{equation}
and for the exciton current
\begin{equation}\label{eq:3001}
J_p=\Gamma_{ext}\rho_{n,n}~.
\end{equation}

The relation between the particle current and heat current is evident from the comparison of equations \ref{eq:16} and \ref{eq:3001}. To plot the currents the steady state solution of Eq. ~\ref{eq:5001} is placed into Eq. ~\ref{eq:3001}. Further details on the calculation are provided in the SI.

\begin{figure}[h]
\centering
\includegraphics[width=1\linewidth]{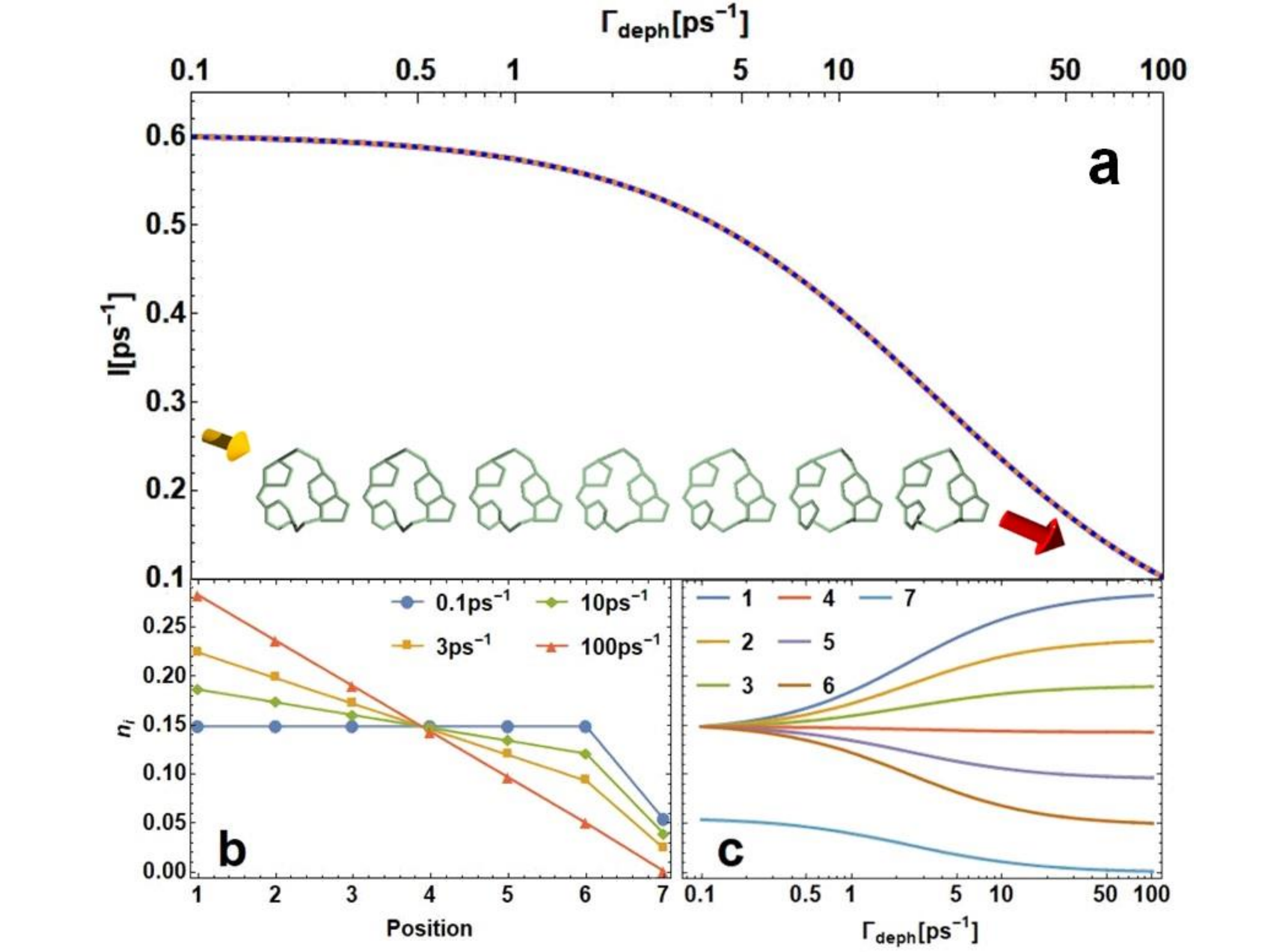}
% \begin{subfigure}{0.8\textwidth}
% \begin{center}
% \includegraphics[width=0.9\linewidth]{fig1Tot.png}
% \end{center}
% \end{subfigure}
% \begin{subfigure}{0.48\textwidth}
%   \includegraphics[width=0.9\linewidth ]{fig1occupSite.jpg} 
%  \end{subfigure}\hfill
%  \begin{subfigure}{0.48\textwidth}
%   \includegraphics[width=0.9\linewidth ]{fig1occupDep.jpg} 
%  \end{subfigure}
%\includegraphics[width=0.7\textwidth,scale=0.3]{fig1occupSite.jpg}
%\includegraphics[width=0.7\textwidth,scale=0.3]{fig1occupDep.jpg}
%\doublespacing
\caption{The linear symmetric chain: (a) Exciton current as a function of dephasing rate. inset: Schematic description of a symmetric 7-sites-chain with uniform energies and distances. The yellow arrow points to the injection site, and the red arrow marks the extraction site. The dashed line is $\Delta_n$ (see Eq.~). (b) Exciton occupation
as a function of site number,  different colors mark different dephasing rates (see legend). (c) Exciton occupations  as a function of dephasing rate, for 7-sites chain. Different colors mark different sites along the chain (see legend).}
\label{fig:SimplestSetup}
\end{figure}

%\subsection{Linear symmetric chain}
We begin our analysis with the simplest symmetric system at hand, namely a uniform chain of \textit{n}-sites and equal on-site-energies, $\epsilon$. Each site interacts with its neighbors via a constant hopping element \textit{t}, as described by the Hamiltonian
\begin{equation}\label{eq:Tb_Of_Chain2}
H=\epsilon\sum_{i=1}^{n} a_i^{\dagger}a_i-t\sum_{i=1}^{n-1} a_i^{\dagger}a_{i+1}+h.c.~.
\end{equation}
The symmetry of the system is reflected not only in the uniformity of the Hamiltonian, but also by the symmetry of the source and sink. In the case of the linear chain, we require an inversion  symmetry between the positions of the source and sink, and place them at the edges of the chain, as depicted in the inset of figure~\ref{fig:SimplestSetup}a: the excitation takes place at the first site (yellow arrow), travels through the chain, and is extracted from the last site (red arrow). 
%\begin{widetext}\begin{center}

%\end{center} \end{widetext}
Figure ~\ref{fig:SimplestSetup}a shows the exciton current for this system, as a function of the dephasing rate $\Gamma_{deph}$. Injection and extraction rates were set to 
$\Gamma_{inj}=\Gamma_{ext}=5$ ps$^{-1}$ , on-site energies $\epsilon=1.23 \times 10^4$ cm$^{-1}$ and coupling elements $t=60 $ cm$^{-1}$, in line with the realistic parameters estimated for the Fenna-Mathiew-Olson (FMO) ETC \cite{Cho2005,Lloyd2011,dubiinterplay2015}.   
As could be expected for the linear symmetric chain \cite{Kassal2012}, increasing the rate of dephasing decreases the current.
%For this system a hand-waving explanation can be given: when the dephasing rate is very low, the %system is fully quantum and uniform, and hence transport can be thought as ballistic (compare to %unitary transmission in electronic systems), implying maximal possible current. Increasing the %dephasing rate introduces classical effects to the system that interrupt the quantum behavior %(analogous to scattering in an electronic system), leading to reduction in current. \\

The analytic relation between exciton current and occupation, Eq.~\ref{eq:3001}, is  motivation to examine exciton occupations, a quantity rarely addressed in the literature. Figure \ref{fig:SimplestSetup}b shows the occupations of the sites (in a linear chain of 7 sites) as a function of site number, for different values of the dephasing rate. In the limit of small dephasing rate (blue spheres), the chain is essentially equally occupied by excitons (with the exception of the extraction site, see SI), reflecting the ballistic nature of the system. With increasing dephasing rate, a density \emph{gradient} gradually forms, with large density at the injection (first) and low density extraction (last) sites. This gradient is most apparent in the fully classical limit of strong dephasing (red triangles). The appearance of a density gradient in the presence of current is a manifestation of Fick's law which relates current to density gradient  \cite{Meixner1965,Kubo1966}. In fact, one can {\sl define} the classical regime as the regime in which the density gradient is fully developed. Note that the occupation of extraction site (site number 7) decreases while the
 gradient is built, and accordingly, the exciton current decreases.
Figure ~\ref{fig:SimplestSetup}c shows the occupations of all sites (each color represent different site number) as a function of dephasing rate; they are uniform in the quantum limit, and spread as the dephasing rate increases. 

The formation of a gradient can be understood from looking at the analytic expressions of the occupations of an $L$-site chain with a side-to-side transport, 
(derivation is detailed in the SI):
%\doublespacing
%\begin{widetext}
\begin{eqnarray}\label{eq:rhoexact}
n_i&=&\frac{m_i}{\sum_i m_i+\frac{\Gamma_{ext}}{\Gamma_{inj}}m_L}~,\nonumber \\
m_i&=&4t^2+(2(L-i)\Gamma_{deph}\Gamma_{ext}+\Gamma_{ext}^2)(1-\delta_{i,L})~~,
\end{eqnarray}
% \end{widetext}
where $n_i$ is the occupation of site $i$ (the diagonal element of the density-matrix, $\rho_{ii}$) in a chain of $L$ sites, and  $\Gamma_{deph},\Gamma_{ext},\Gamma_{inj}$ are the rate of dephasing, extraction and injection, respectively. The expressions in Eq.~ \ref{eq:rhoexact} reveals the effect of dephasing on the exciton density distribution in the chain; the second expression of $m_i$ shows the formation of the linear slope, with a gradient which is proportional to $\Gamma_{deph}$. Furthermore, it can be seen that the more distant the site is to the extraction point, the more pronounced will the effect of dephasing be on the occupation. It is clear from Eq.~\ref{eq:rhoexact} that the density at the extraction site is monotonously decreasing, leading to a monotonic decrease in exciton current.
  
\begin{figure}
\centering
% \begin{subfigure}{0.8\textwidth}
% \centering
% \includegraphics[width=0.75\linewidth]{fig2ParticlePic.jpg}
% \centering
% \end{subfigure}
% \begin{subfigure}{0.48\textwidth}
%   \includegraphics[width=0.9\linewidth ]{fig2occupSite.jpg} 
%  \end{subfigure}\hfill
%  \begin{subfigure}{0.48\textwidth}
%   \includegraphics[width=0.9\linewidth ]{fig2occupDep.jpg} 
%  \end{subfigure}
%\includegraphics[width=0.7\textwidth,scale=0.3]{fig1occupSite.jpg}
\includegraphics[width=1\linewidth]{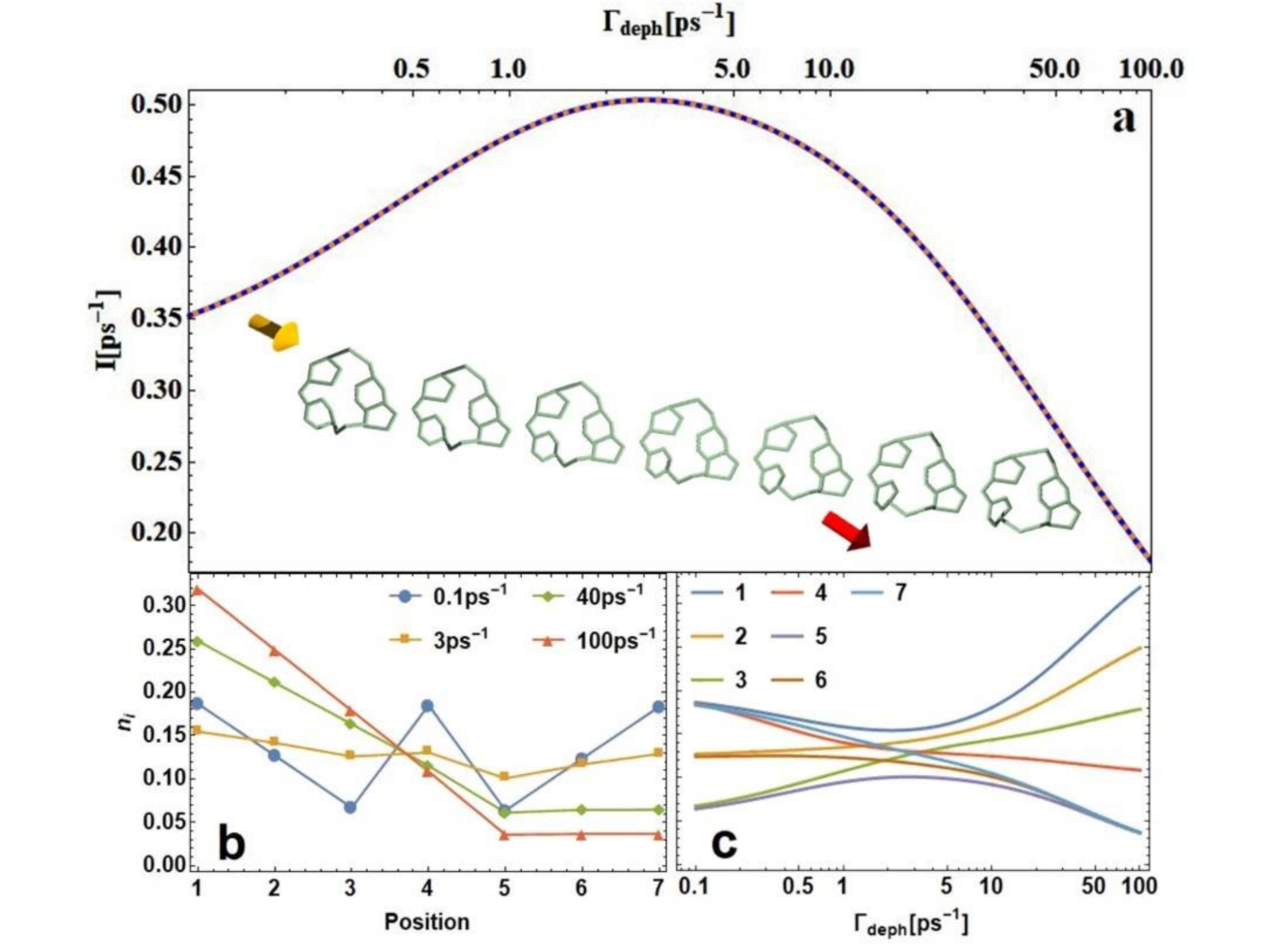}
%\doublespacing
\caption{The linear non-symmetric chain: (a)-(c) Same as in Fig.~\ref{fig:SimplestSetup}, for the 7-site linear non-symmetric chain. The non-symmetric chain shows a \emph{qualitative} difference compared to the symmetric chain, manifested through a non-monotonic dependence of the current on dephasing rate, signaling the appearance of ENAQT.} 
\label{fig:middlextract}
\end{figure}

%\subsection{The non-symmetric linear chain}
So far as the symmetric linear chain  is considered, no ENAQT is observed. However, it appears upon a slight modification of the system \cite{Kassal2012}.   
Consider the same uniform linear chain only with a slight change: the extraction site is moved away from the edge of the chain (thus breaking the inversion symmetry), schematically described in the inset of figure ~\ref{fig:middlextract}a. Surprisingly, this seemingly minor difference yields qualitatively different behavior. The exciton current as a function of dephasing (main figure ~\ref{fig:middlextract}a) displays a non-monotonic dependence, with a maximum in the current at a finite $\Gamma_{deph}$\cite{Kassal2012}, signalling the appearances of ENAQT. We stress here that dephasing is a dissipative process, and yet the exction current (and consequently the energy flow) are enhanced in its presence. 
%\begin{widetext}\begin{center}
%\end{center}\end{widetext}

Figure ~\ref{fig:middlextract}b shows the exciton occupation along a (non-symmetric) 7-site chain as a function of site number, for different dephasing rates. In contrast with the symmetric setup, here the occupations are not uniform in the quantum limit (blue spheres), reflecting the structure of the wave-functions and their interplay with the source and sink. With the increase in the dephasing rate, a uniform density gradient is formed between the injection and extraction sites. While the gradient is formed,
the occupation of the extraction site increases and then decreases, and since the exciton current is proportional to the extraction site occupation, 
it acts similarly.

This can be seen more clearly in Figure ~\ref{fig:middlextract}c, which shows the exciton occupations as a function of dephasing rate.  We observe that the transition from a wide distribution of occupations (at $\Gamma_{deph}=0$) to a linear gradient ($\Gamma_{deph}=100 $ps$^{-1}$) passes through an intermediate stage  where occupations along the chain become similar (at $\Gamma_{deph}\approx5 $ps$^{-1}$). This behavior is a result of two competing processes. The first is the direct outcome of dephasing, which can be considered the result of an instantaneous "measurement" of the system by the environment at a random site \cite{breuer2002theory}, implying a full mixing of the system eigen-states. As a result, the real-space occupations tend to average into a narrower distribution \cite{PhysRevLett.85.812} (see SI for an example). The second process is the formation of the density gradient. While the gradient shape is determined by the positions of the  extraction and the injection sites (see figure ~\ref{fig:SimplestSetup}b and ~\ref{fig:middlextract}b), its formation is enabled by the dephasing process. This can be deduced from the dependence of $m_i$ in equation ~\ref{eq:rhoexact} on the dephasing rate. For a small dephasing rate, the position-dependent term (which is responsible for the gradient) is  small compared with the first, position-independent part, while for large dephasing rate it is the dominant factor in determining $n_i$. The crossover between these two regimes leads to the non-monotonic shape seen in figure  ~\ref{fig:middlextract}c.

%\subsection{Correlation between exciton current and density}
Comparing Figure ~\ref{fig:middlextract}c with Figure ~\ref{fig:middlextract}a reveals  a correlation between the exciton current and the distribution of the occupations:  the maximal current seems to appear at (or close to) the dephasing rate at which the spread of occupations is minimal. 
To quantify this relation between the distribution of exciton occupations and optimal current, we define 
the quantity 
 
\begin{equation}\label{eq:91409}
\Delta_{n}=1-\sqrt{\sum_i \bigg(\langle n_i\rangle - n_{ext}\bigg) ^2}~, 
\end{equation}

where $n_{ext}$ is the occupation of the extraction site, and $\langle n_i \rangle$ is the average occupation of the $i$-th site. $\Delta_{n}$ is a measure of the spread of the occupations, and as such, should exhibit a maximum at the same dephasing rate where the spread is minimal and current is maximal. In fact, this relation can be derived analytically within Lindblad theory under certain limitations (see SI). In Figs.~ \ref{fig:SimplestSetup}a and \ref{fig:middlextract}a, $\Delta_n$ is plotted (in arbitrary units) on top of the current (dashed lines). One can clearly see how the behavior of $\Delta_n$ follows that of the exciton current.  

\begin{figure}
\centering
\includegraphics[width=1\textwidth,scale=0.3]{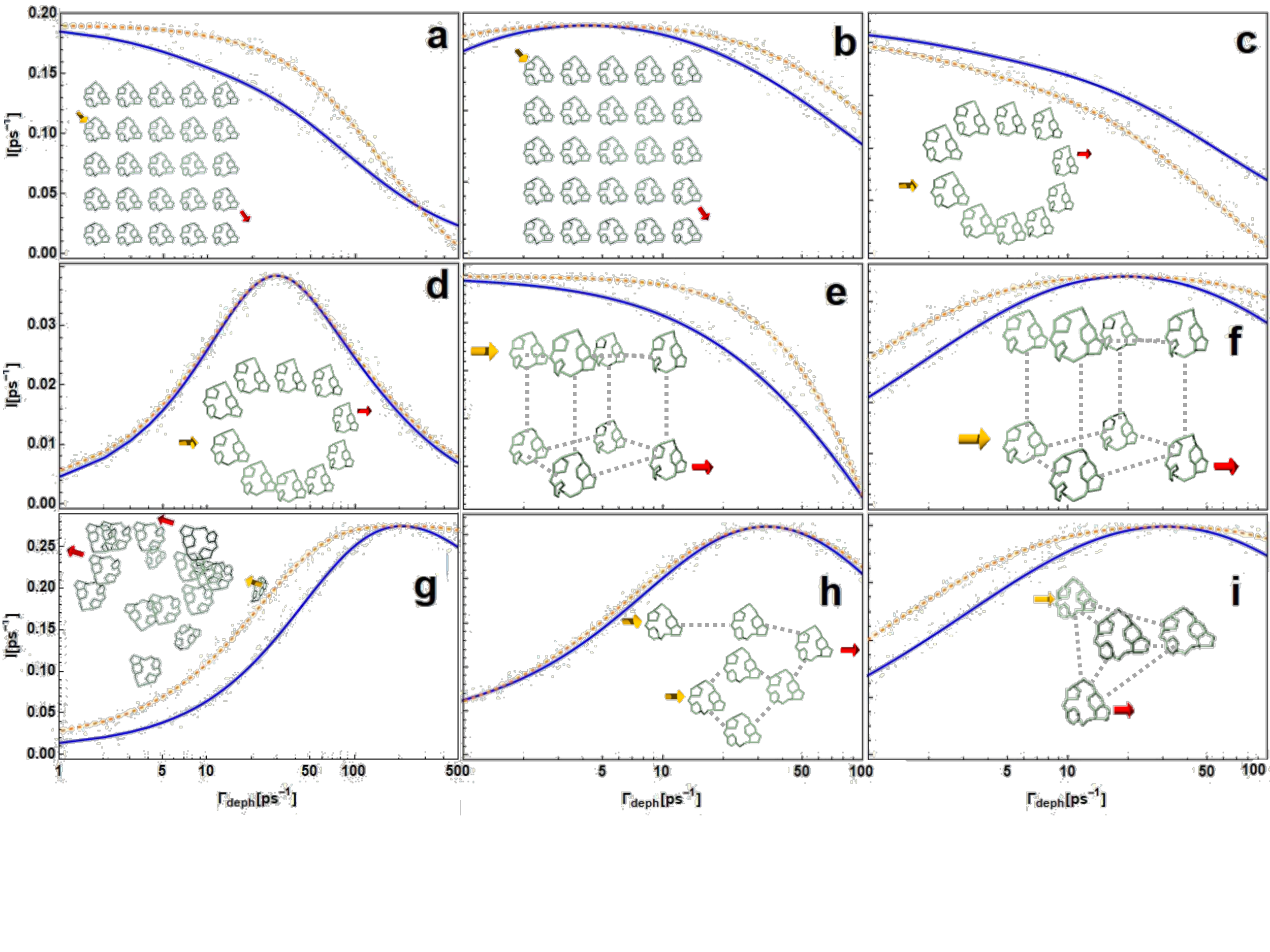}
\caption{Correlation between exciton current and density: Exciton current (Blue lines) and $\Delta_n$ (Orange dashed lines) as a function of dephasing rate for different network geometries and topologies:\textbf{(a)} A $5 \times 5$  network of chromophores with inversion symmetry, \textbf{b)} A $5 \times 5$ network of chromophores without inversion symmetry, \textbf{(c)} a ring of chromophores with uniform energies, \textbf{(d)} A ring of chromophores with random energies ($10^2-10^5$ cm$^{-1}$), \textbf{(e)} Cube of chromophores in a symmetrical setup, \textbf{(f)} Cube of chromophores in a non-symmetric setup, \textbf{(g)} full-graph of 16 chromophores with random energies and distances, the energy is extracted from two extraction-sites, \textbf{(h)} Biological setup-the FMO exciton network, \textbf{(i)} A pyramid-like network of chromophores. Two features are notable: 1) Non-monotonic behavior of the current appears only when there is no inversion symmetry, 2) in all examples, $\Delta_n$ follows the same behavior as the exciton current. }
\label{fig:Examples}
\end{figure}

To demonstrate the universality of this relation, we examine it in larger and more complex networks. 
Figure ~\ref{fig:Examples} shows the exciton current $I$ (blue line) and $\Delta_{n}$ (dashed orange line) as a function of dephasing rate, for selected networks of different topologies, dimensions, sizes and symmetries. 
 As seen, the two quantities closely follow each other, and if there is a maximum in the particle current, $\Delta_{n}$ also exhibits a maximum, and at the same dephasing rate.
%\begin{widetext}\begin{center}
%\end{center}
The results shown insofar were obtained by evaluating the steady-state solution of the Lindblad equation. We argue that these results do not depend on the calculation method. To show this we have calculated the currents as a function of dephasing rate for the same system (i.e. the symmetric and non-symmetric chains and the examples of Fig. ~3) in two additional methods. The first is the full Redfield equation, which takes into account the spectral properties of the environment. The second is the time-dependent Lindblad equation, where a pulse-excitation was considered, and the current as a function of time was evaluated (and integrated to obtain the total current). In both cases we found the same results, namely that non-symmetric networks exhibit ENAQT, and that the behavior of the current correlates with $\Delta_n$, thus supporting our claims (details and results of these calculations are in the SI). We conjecture (and leave the verification to future studies) that these features persist beyond the Markovian limit, as steady-state currents should only be weakly affected by non-Markovianity \cite{diosi1997,rebentrost2009non, Dutta2017}. 

The question still remains, why does ENAQT only appear in non-symmetric networks, which do not posses an inversion symmetry. In the presence of an inversion symmetry (which includes, as mentioned above, interchanging the source and drain terms), the master equation for inversion points (points connected by inversion, except for the source and drain sites) are exactly the same. It follows that the density matrix itself is symmetric under inversion and every two inversion-related sites will have the same density, thus reducing the density fluctuations. In this case, the dephasing works only to form the density gradient, leading to a monotonic decrease in the sink site density and, respectively, the current. Put simply, ENAQT only occurs if the exciton density is non-uniform in the fully quantum limit, which is never the case in a uniform system with an inversion symmetry. 

%We point that our results pertain to the presence of ENAQT of a single exciton in a single exciton transfer complex. It is of interest to find out whether our results hold generally for many-exciton systems \cite{gurvitz2017multiscale} and coupled ETCs \cite{mendoza2014transport}, as well as to other systems where dephasing may play an important role, e.g. electronic transport through bio-molecules and molecular junctions \cite{penazzi2016self,kocherzhenko2010charge,nozaki2012disorder,contreras2014dephasing,xiang2015intermediate}.

Comparing between figures \ref{fig:middlextract}(a) and \ref{fig:SimplestSetup}(a) one can see that the exciton current is actually higher in the coherent regime for the symmetric system vs the asymmetric system. However, if disorder, asymmetry or a dephasing environment are unavoidably present (which seems to be the case for natural photo-synthetic complexes), the intermediate coherent-dephasing regime delivers better performance. Comparing the transport properties of different geometries can thus be an important tool for understanding transport mechanisms in artificial ETCs 
\cite{Eisenberg2014,eisenberg2017concentration,banal2017photophysics,boulais2017programmed},
as well as in other systems where dephasing may play an important role, e.g. electronic transport through bio-molecules and molecular junctions \cite{penazzi2016self,kocherzhenko2010charge,nozaki2012disorder,contreras2014dephasing,xiang2015intermediate}.

\begin{acknowledgement}
We acknowledge support from the Adelis foundation. E.Z-.H. Acknowledges support from the IKI interdisciplinary fellowship. 
\end{acknowledgement}

% The Lindblad calculations are supplemented by the
% Redfield equation, given by \cite{Breuer2002}:
% \begin{equation}\begin{aligned}\label{eq:215}
% 0=-i[H_s,\rho]+
% \sum_{\omega}\sum_{\alpha,\beta}\Gamma_{\alpha\beta}(\omega)
% (A_\alpha(\omega)\rho_s(t)A_\beta(\omega)^{\dagger}\\-A_\beta(\omega)^{\dagger} A_\alpha(\omega)\rho_s(t))+h.c
% \end{aligned}\end{equation}
% $H_s$ is the system Hamiltonian and the operators $A_\alpha(\omega)$, $A^\dagger_\alpha(\omega)$, $A_\beta(\omega)$ and $A^\dagger_\beta(\omega)$ 
%  are the eigenoperators of $Hs$ belonging to the frequencies $\omega$, respectively. The frequencies are defied as $\omega=E_{\alpha}-E_{\beta}$.
% $\Gamma_{\alpha\beta}(\omega)$ describes the interaction of the system with the environment,at steady state it can be described as  \cite{purkayasth2016}
% \begin{equation}
% \Gamma_{\alpha\beta}(\omega)=n(\omega)J(\omega)
% \end{equation}
% And $\Gamma_{\alpha\beta}$ is the ohmic spectral function \cite{Pachon2012}\\

\begin{suppinfo} Detailed methods description. Relation between exciton occupations and current. Derivation of occupations in a linear symmetric chain. Effect of dephasing on occupations in the absence of current. Calculation using the Redfield equation. Detailed examples of exciton occupation as a function of dephasing rate. Calculation using a thermodynamically-consistent Lindblad equation. 
\end{suppinfo}

\bibliographystyle{unsrt}
%\bibliography{ref3.bib}

\begin{mcitethebibliography}{60}
\providecommand*\natexlab[1]{#1}
\providecommand*\mciteSetBstSublistMode[1]{}
\providecommand*\mciteSetBstMaxWidthForm[2]{}
\providecommand*\mciteBstWouldAddEndPuncttrue
  {\def\EndOfBibitem{\unskip.}}
\providecommand*\mciteBstWouldAddEndPunctfalse
  {\let\EndOfBibitem\relax}
\providecommand*\mciteSetBstMidEndSepPunct[3]{}
\providecommand*\mciteSetBstSublistLabelBeginEnd[3]{}
\providecommand*\EndOfBibitem{}
\mciteSetBstSublistMode{f}
\mciteSetBstMaxWidthForm{subitem}{(\alph{mcitesubitemcount})}
\mciteSetBstSublistLabelBeginEnd
  {\mcitemaxwidthsubitemform\space}
  {\relax}
  {\relax}

\bibitem[Mohseni \latin{et~al.}(2014)Mohseni, Omar, Engel, and
  Plenio]{mohseni2014quantum}
Mohseni,~M.; Omar,~Y.; Engel,~G.~S.; Plenio,~M.~B. \emph{Quantum effects in
  biology}; Cambridge University Press, 2014\relax
\mciteBstWouldAddEndPuncttrue
\mciteSetBstMidEndSepPunct{\mcitedefaultmidpunct}
{\mcitedefaultendpunct}{\mcitedefaultseppunct}\relax
\EndOfBibitem
\bibitem[Govorov \latin{et~al.}(2016)Govorov, MartÃ­nez, and
  Demir]{govorov2016}
Govorov,~A.; MartÃ­nez,~P. L.~H.; Demir,~H.~V. \emph{Understanding and
  Modeling F\"{o}rster type Resonance Energy Transfer (FRET) introduction to
  FRET}; Springer, 2016\relax
\mciteBstWouldAddEndPuncttrue
\mciteSetBstMidEndSepPunct{\mcitedefaultmidpunct}
{\mcitedefaultendpunct}{\mcitedefaultseppunct}\relax
\EndOfBibitem
\bibitem[Levi \latin{et~al.}(2015)Levi, Mostarda, Rao, and Mintert]{levi2015}
Levi,~F.; Mostarda,~S.; Rao,~F.; Mintert,~F. \emph{Rep. Prog. Phys.}
  \textbf{2015}, \emph{78}, 082001\relax
\mciteBstWouldAddEndPuncttrue
\mciteSetBstMidEndSepPunct{\mcitedefaultmidpunct}
{\mcitedefaultendpunct}{\mcitedefaultseppunct}\relax
\EndOfBibitem
\bibitem[Fleming and Scholes(2004)Fleming, and Scholes]{fleming2004}
Fleming,~G.~R.; Scholes,~G.~D. \emph{Nature} \textbf{2004}, \emph{431},
  256--257\relax
\mciteBstWouldAddEndPuncttrue
\mciteSetBstMidEndSepPunct{\mcitedefaultmidpunct}
{\mcitedefaultendpunct}{\mcitedefaultseppunct}\relax
\EndOfBibitem
\bibitem[Engel \latin{et~al.}(2007)Engel, Calhoun, Read, Ahn, ManÄal, Cheng,
  Blankenship, and Fleming]{engel2007}
Engel,~G.~S.; Calhoun,~T.~R.; Read,~E.~L.; Ahn,~T.-K.; ManÄal,~T.;
  Cheng,~Y.-C.; Blankenship,~R.~E.; Fleming,~G.~R. \emph{Nature} \textbf{2007},
  \emph{446}, 782--786\relax
\mciteBstWouldAddEndPuncttrue
\mciteSetBstMidEndSepPunct{\mcitedefaultmidpunct}
{\mcitedefaultendpunct}{\mcitedefaultseppunct}\relax
\EndOfBibitem
\bibitem[Calhoun \latin{et~al.}(2009)Calhoun, Ginsberg, Schlau-Cohen, Cheng,
  Ballottari, Bassi, and Fleming]{Calhoun2009}
Calhoun,~T.~R.; Ginsberg,~N.~S.; Schlau-Cohen,~G.~S.; Cheng,~Y.-C.;
  Ballottari,~M.; Bassi,~R.; Fleming,~G.~R. \emph{The Journal of Physical
  Chemistry B} \textbf{2009}, \emph{113}, 16291--16295\relax
\mciteBstWouldAddEndPuncttrue
\mciteSetBstMidEndSepPunct{\mcitedefaultmidpunct}
{\mcitedefaultendpunct}{\mcitedefaultseppunct}\relax
\EndOfBibitem
\bibitem[Collini \latin{et~al.}(2010)Collini, Wong, Wilk, Curmi, Brumer, and
  Scholes]{Collini2010}
Collini,~E.; Wong,~C.~Y.; Wilk,~K.~E.; Curmi,~P.~M.; Brumer,~P.; Scholes,~G.~D.
  \emph{Nature} \textbf{2010}, \emph{463}, 644--647\relax
\mciteBstWouldAddEndPuncttrue
\mciteSetBstMidEndSepPunct{\mcitedefaultmidpunct}
{\mcitedefaultendpunct}{\mcitedefaultseppunct}\relax
\EndOfBibitem
\bibitem[Panitchayangkoon \latin{et~al.}(2011)Panitchayangkoon, Voronine,
  Abramavicius, Caram, Lewis, Mukamel, and Engel]{Panitchayangkoon2011}
Panitchayangkoon,~G.; Voronine,~D.~V.; Abramavicius,~D.; Caram,~J.~R.;
  Lewis,~N.~H.; Mukamel,~S.; Engel,~G.~S. \emph{Proceedings of the National
  Academy of Sciences} \textbf{2011}, \emph{108}, 20908--20912\relax
\mciteBstWouldAddEndPuncttrue
\mciteSetBstMidEndSepPunct{\mcitedefaultmidpunct}
{\mcitedefaultendpunct}{\mcitedefaultseppunct}\relax
\EndOfBibitem
\bibitem[Ishizaki and Fleming(2012)Ishizaki, and Fleming]{Ishizaki2012}
Ishizaki,~A.; Fleming,~G.~R. \emph{Annu. Rev. Condens. Matter Phys.}
  \textbf{2012}, \emph{3}, 333--361\relax
\mciteBstWouldAddEndPuncttrue
\mciteSetBstMidEndSepPunct{\mcitedefaultmidpunct}
{\mcitedefaultendpunct}{\mcitedefaultseppunct}\relax
\EndOfBibitem
\bibitem[Collini(2013)]{Collini2013}
Collini,~E. \emph{Chemical Society Reviews} \textbf{2013}, \emph{42},
  4932--4947\relax
\mciteBstWouldAddEndPuncttrue
\mciteSetBstMidEndSepPunct{\mcitedefaultmidpunct}
{\mcitedefaultendpunct}{\mcitedefaultseppunct}\relax
\EndOfBibitem
\bibitem[Lambert \latin{et~al.}(2013)Lambert, Chen, Cheng, Li, Chen, and
  Nori]{Lambert2013}
Lambert,~N.; Chen,~Y.-N.; Cheng,~Y.-C.; Li,~C.-M.; Chen,~G.-Y.; Nori,~F.
  \emph{Nature Physics} \textbf{2013}, \emph{9}, 10--18\relax
\mciteBstWouldAddEndPuncttrue
\mciteSetBstMidEndSepPunct{\mcitedefaultmidpunct}
{\mcitedefaultendpunct}{\mcitedefaultseppunct}\relax
\EndOfBibitem
\bibitem[Pach{\'o}n and Brumer(2012)Pach{\'o}n, and Brumer]{Pachon2012}
Pach{\'o}n,~L.~A.; Brumer,~P. \emph{Physical Chemistry Chemical Physics}
  \textbf{2012}, \emph{14}, 10094--10108\relax
\mciteBstWouldAddEndPuncttrue
\mciteSetBstMidEndSepPunct{\mcitedefaultmidpunct}
{\mcitedefaultendpunct}{\mcitedefaultseppunct}\relax
\EndOfBibitem
\bibitem[Kassal \latin{et~al.}(2013)Kassal, Yuen-Zhou, and
  Rahimi-Keshari]{kassal2013}
Kassal,~I.; Yuen-Zhou,~J.; Rahimi-Keshari,~S. \emph{J. Phys. Chem. Lett.}
  \textbf{2013}, \emph{4}, 362--367\relax
\mciteBstWouldAddEndPuncttrue
\mciteSetBstMidEndSepPunct{\mcitedefaultmidpunct}
{\mcitedefaultendpunct}{\mcitedefaultseppunct}\relax
\EndOfBibitem
\bibitem[Miller(2012)]{Miller2012}
Miller,~W.~H. \emph{The Journal of chemical physics} \textbf{2012}, \emph{136},
  210901\relax
\mciteBstWouldAddEndPuncttrue
\mciteSetBstMidEndSepPunct{\mcitedefaultmidpunct}
{\mcitedefaultendpunct}{\mcitedefaultseppunct}\relax
\EndOfBibitem
\bibitem[Ritschel \latin{et~al.}(2011)Ritschel, Roden, Strunz, Aspuru-Guzik,
  and Eisfeld]{ritschel2011}
Ritschel,~G.; Roden,~J.; Strunz,~W.~T.; Aspuru-Guzik,~A.; Eisfeld,~A. \emph{J.
  Phys. Chem. Lett.} \textbf{2011}, \emph{2}, 2912--2917\relax
\mciteBstWouldAddEndPuncttrue
\mciteSetBstMidEndSepPunct{\mcitedefaultmidpunct}
{\mcitedefaultendpunct}{\mcitedefaultseppunct}\relax
\EndOfBibitem
\bibitem[Duan \latin{et~al.}(2017)Duan, Prokhorenko, Cogdell, Ashraf, Stevens,
  Thorwart, and Miller]{duan2017nature}
Duan,~H.-G.; Prokhorenko,~V.~I.; Cogdell,~R.~J.; Ashraf,~K.; Stevens,~A.~L.;
  Thorwart,~M.; Miller,~R.~D. \emph{Proceedings of the National Academy of
  Sciences} \textbf{2017}, \emph{114}, 8493--8498\relax
\mciteBstWouldAddEndPuncttrue
\mciteSetBstMidEndSepPunct{\mcitedefaultmidpunct}
{\mcitedefaultendpunct}{\mcitedefaultseppunct}\relax
\EndOfBibitem
\bibitem[Semiao \latin{et~al.}(2010)Semiao, Furuya, and Milburn]{Semiao2010}
Semiao,~F.~L.; Furuya,~K.; Milburn,~G.~J. \emph{New Journal of Physics}
  \textbf{2010}, \emph{12}, 083033\relax
\mciteBstWouldAddEndPuncttrue
\mciteSetBstMidEndSepPunct{\mcitedefaultmidpunct}
{\mcitedefaultendpunct}{\mcitedefaultseppunct}\relax
\EndOfBibitem
\bibitem[Nalbach \latin{et~al.}(2010)Nalbach, Eckel, and Thorwart]{Nalbach2010}
Nalbach,~P.; Eckel,~J.; Thorwart,~M. \emph{New Journal of Physics}
  \textbf{2010}, \emph{12}, 065043\relax
\mciteBstWouldAddEndPuncttrue
\mciteSetBstMidEndSepPunct{\mcitedefaultmidpunct}
{\mcitedefaultendpunct}{\mcitedefaultseppunct}\relax
\EndOfBibitem
\bibitem[Scholak \latin{et~al.}(2011)Scholak, de~Melo, Wellens, Mintert, and
  Buchleitner]{Scholak2011a}
Scholak,~T.; de~Melo,~F.; Wellens,~T.; Mintert,~F.; Buchleitner,~A. \emph{Phys.
  Rev. E} \textbf{2011}, \emph{83}, 021912\relax
\mciteBstWouldAddEndPuncttrue
\mciteSetBstMidEndSepPunct{\mcitedefaultmidpunct}
{\mcitedefaultendpunct}{\mcitedefaultseppunct}\relax
\EndOfBibitem
\bibitem[Ajisaka \latin{et~al.}(2015)Ajisaka, Zunkovic, and Dubi]{Ajisaka2015}
Ajisaka,~S.; Zunkovic,~B.; Dubi,~Y. \emph{Sci. Rep.} \textbf{2015}, \emph{5},
  8312\relax
\mciteBstWouldAddEndPuncttrue
\mciteSetBstMidEndSepPunct{\mcitedefaultmidpunct}
{\mcitedefaultendpunct}{\mcitedefaultseppunct}\relax
\EndOfBibitem
\bibitem[Lim \latin{et~al.}(2014)Lim, Tame, Yee, Lee, and Lee]{Lim2014}
Lim,~J.; Tame,~M.; Yee,~K.~H.; Lee,~J.-S.; Lee,~J. \emph{New Journal of
  Physics} \textbf{2014}, \emph{16}, 053018\relax
\mciteBstWouldAddEndPuncttrue
\mciteSetBstMidEndSepPunct{\mcitedefaultmidpunct}
{\mcitedefaultendpunct}{\mcitedefaultseppunct}\relax
\EndOfBibitem
\bibitem[Le{\'o}n-Montiel \latin{et~al.}(2015)Le{\'o}n-Montiel,
  Quiroz-Ju{\'a}rez, Quintero-Torres, Dom{\'\i}nguez-Ju{\'a}rez, Moya-Cessa,
  Torres, and Arag{\'o}n]{leon2015noise}
Le{\'o}n-Montiel,~R. d.~J.; Quiroz-Ju{\'a}rez,~M.~A.; Quintero-Torres,~R.;
  Dom{\'\i}nguez-Ju{\'a}rez,~J.~L.; Moya-Cessa,~H.~M.; Torres,~J.~P.;
  Arag{\'o}n,~J.~L. \emph{Scientific reports} \textbf{2015}, \emph{5}\relax
\mciteBstWouldAddEndPuncttrue
\mciteSetBstMidEndSepPunct{\mcitedefaultmidpunct}
{\mcitedefaultendpunct}{\mcitedefaultseppunct}\relax
\EndOfBibitem
\bibitem[Biggerstaff \latin{et~al.}(2016)Biggerstaff, Heilmann, Zecevik,
  Gr{\"a}fe, Broome, Fedrizzi, Nolte, Szameit, White, and
  Kassal]{Biggerstaff2016}
Biggerstaff,~D.~N.; Heilmann,~R.; Zecevik,~A.~A.; Gr{\"a}fe,~M.; Broome,~M.~A.;
  Fedrizzi,~A.; Nolte,~S.; Szameit,~A.; White,~A.~G.; Kassal,~I. \emph{Nature
  communications} \textbf{2016}, \emph{7}\relax
\mciteBstWouldAddEndPuncttrue
\mciteSetBstMidEndSepPunct{\mcitedefaultmidpunct}
{\mcitedefaultendpunct}{\mcitedefaultseppunct}\relax
\EndOfBibitem
\bibitem[Viciani \latin{et~al.}(2016)Viciani, Gherardini, Lima, Bellini, and
  Caruso]{viciani2016disorder}
Viciani,~S.; Gherardini,~S.; Lima,~M.; Bellini,~M.; Caruso,~F. \emph{Scientific
  reports} \textbf{2016}, \emph{6}, 37791\relax
\mciteBstWouldAddEndPuncttrue
\mciteSetBstMidEndSepPunct{\mcitedefaultmidpunct}
{\mcitedefaultendpunct}{\mcitedefaultseppunct}\relax
\EndOfBibitem

\bibitem[Caruso \latin{et~al.}(2016)Caruso, Crespi, Ciriolo, Sciarrino, and
  Osellame]{Caruso2016}
Caruso,~F.;Crespi,~A.; Ciriolo,~A. G.; Sciarrino,~F.; Osellame,~R. \emph{Nature Communications} \textbf{2016}, \emph{7}, 11682\relax
\mciteBstWouldAddEndPuncttrue
\mciteSetBstMidEndSepPunct{\mcitedefaultmidpunct}
{\mcitedefaultendpunct}{\mcitedefaultseppunct}\relax
\EndOfBibitem


\bibitem[Rebentrost \latin{et~al.}(2009)Rebentrost, Mohseni, Kassal, Lloyd, and
  Aspuru-Guzik]{Rebentrost2009}
Rebentrost,~P.; Mohseni,~M.; Kassal,~I.; Lloyd,~S.; Aspuru-Guzik,~A. \emph{New
  J. Phys.} \textbf{2009}, \emph{11}, 033003\relax
\mciteBstWouldAddEndPuncttrue
\mciteSetBstMidEndSepPunct{\mcitedefaultmidpunct}
{\mcitedefaultendpunct}{\mcitedefaultseppunct}\relax
\EndOfBibitem
\bibitem[Manzano(2013)]{Manzano2013}
Manzano,~D. \emph{PLOS ONE} \textbf{2013}, \emph{8}\relax
\mciteBstWouldAddEndPuncttrue
\mciteSetBstMidEndSepPunct{\mcitedefaultmidpunct}
{\mcitedefaultendpunct}{\mcitedefaultseppunct}\relax
\EndOfBibitem
\bibitem[Chin \latin{et~al.}(2010)Chin, Datta, Caruso, Huelga, and
  Plenio]{Chin2010}
Chin,~A.~W.; Datta,~A.; Caruso,~F.; Huelga,~S.~F.; Plenio,~M.~B. \emph{New J.
  Phys.} \textbf{2010}, \emph{12}, 065002\relax
\mciteBstWouldAddEndPuncttrue
\mciteSetBstMidEndSepPunct{\mcitedefaultmidpunct}
{\mcitedefaultendpunct}{\mcitedefaultseppunct}\relax
\EndOfBibitem
\bibitem[Cao and Silbey(2009)Cao, and Silbey]{cao2009optimization}
Cao,~J.; Silbey,~R.~J. \emph{J. Phys. Chem. A} \textbf{2009}, \emph{113},
  13825--13838\relax
\mciteBstWouldAddEndPuncttrue
\mciteSetBstMidEndSepPunct{\mcitedefaultmidpunct}
{\mcitedefaultendpunct}{\mcitedefaultseppunct}\relax
\EndOfBibitem
\bibitem[Li \latin{et~al.}(2015)Li, Caruso, Gauger, and Benjamin]{Li2015}
Li,~Y.; Caruso,~F.; Gauger,~E.; Benjamin,~S.~C. \emph{New Journal of Physics}
  \textbf{2015}, \emph{17}, 013057\relax
\mciteBstWouldAddEndPuncttrue
\mciteSetBstMidEndSepPunct{\mcitedefaultmidpunct}
{\mcitedefaultendpunct}{\mcitedefaultseppunct}\relax
\EndOfBibitem
\bibitem[Plenio and Huelga(2008)Plenio, and Huelga]{Plenio2008}
Plenio,~M.~B.; Huelga,~S.~F. \emph{New J. Phys.} \textbf{2008}, \emph{10},
  113019\relax
\mciteBstWouldAddEndPuncttrue
\mciteSetBstMidEndSepPunct{\mcitedefaultmidpunct}
{\mcitedefaultendpunct}{\mcitedefaultseppunct}\relax
\EndOfBibitem
\bibitem[Caruso \latin{et~al.}(2009)Caruso, Chin, Datta, Huelga, and
  Plenio]{Caruso2009}
Caruso,~F.; Chin,~A.~W.; Datta,~A.; Huelga,~S.~F.; Plenio,~M.~B. \emph{The
  Journal of Chemical Physics} \textbf{2009}, \emph{131}, 105106\relax
\mciteBstWouldAddEndPuncttrue
\mciteSetBstMidEndSepPunct{\mcitedefaultmidpunct}
{\mcitedefaultendpunct}{\mcitedefaultseppunct}\relax
\EndOfBibitem
\bibitem[Wu \latin{et~al.}(2010)Wu, Liu, Shen, Cao, and
  Silbey]{wu2010efficient}
Wu,~J.; Liu,~F.; Shen,~Y.; Cao,~J.; Silbey,~R.~J. \emph{New Journal of Physics}
  \textbf{2010}, \emph{12}, 105012\relax
\mciteBstWouldAddEndPuncttrue
\mciteSetBstMidEndSepPunct{\mcitedefaultmidpunct}
{\mcitedefaultendpunct}{\mcitedefaultseppunct}\relax
\EndOfBibitem
\bibitem[Nesterov \latin{et~al.}(2013)Nesterov, Berman,
  S{\'a}nchez~Mart{\'i}nez, and Sayre]{Nesterov2013}
Nesterov,~A.~I.; Berman,~G.~P.; S{\'a}nchez~Mart{\'i}nez,~J.~M.; Sayre,~R.~T.
  \emph{Journal of Mathematical Chemistry} \textbf{2013}, \emph{51},
  2514--2541\relax
\mciteBstWouldAddEndPuncttrue
\mciteSetBstMidEndSepPunct{\mcitedefaultmidpunct}
{\mcitedefaultendpunct}{\mcitedefaultseppunct}\relax
\EndOfBibitem
\bibitem[Berman \latin{et~al.}(2015)Berman, Nesterov, Lֳ³pez, and
  Sayre]{Berman2015}
Berman,~G.~P.; Nesterov,~A.~I.; Lֳ³pez,~G.~V.; Sayre,~R.~T. \emph{The Journal
  of Physical Chemistry C} \textbf{2015}, \emph{119}, 22289--22296\relax
\mciteBstWouldAddEndPuncttrue
\mciteSetBstMidEndSepPunct{\mcitedefaultmidpunct}
{\mcitedefaultendpunct}{\mcitedefaultseppunct}\relax
\EndOfBibitem
\bibitem[Baghbanzadeh and Kassal(2016)Baghbanzadeh, and
  Kassal]{Baghbanzadeh2016}
Baghbanzadeh,~S.; Kassal,~I. \emph{Phys. Chem. Chem. Phys.} \textbf{2016},
  \emph{18}, 7459--7467\relax
\mciteBstWouldAddEndPuncttrue
\mciteSetBstMidEndSepPunct{\mcitedefaultmidpunct}
{\mcitedefaultendpunct}{\mcitedefaultseppunct}\relax
\EndOfBibitem
\bibitem[Wu \latin{et~al.}(2013)Wu, Silbey, and Cao]{PhysRevLett.110.200402}
Wu,~J.; Silbey,~R.~J.; Cao,~J. \emph{Phys. Rev. Lett.} \textbf{2013},
  \emph{110}, 200402\relax
\mciteBstWouldAddEndPuncttrue
\mciteSetBstMidEndSepPunct{\mcitedefaultmidpunct}
{\mcitedefaultendpunct}{\mcitedefaultseppunct}\relax
\EndOfBibitem
\bibitem[Dubi(2015)]{dubiinterplay2015}
Dubi,~Y. \emph{J. Phys. Chem. C} \textbf{2015}, \emph{119}, 25252--25259\relax
\mciteBstWouldAddEndPuncttrue
\mciteSetBstMidEndSepPunct{\mcitedefaultmidpunct}
{\mcitedefaultendpunct}{\mcitedefaultseppunct}\relax
\EndOfBibitem
\bibitem[Breuer and Petruccione(2002)Breuer, and Petruccione]{breuer2002theory}
Breuer,~H.-P.; Petruccione,~F. \emph{The theory of open quantum systems};
  Oxford University Press on Demand, 2002\relax
\mciteBstWouldAddEndPuncttrue
\mciteSetBstMidEndSepPunct{\mcitedefaultmidpunct}
{\mcitedefaultendpunct}{\mcitedefaultseppunct}\relax
\EndOfBibitem
\bibitem[Gelbwaser-Klimovsky and Aspuru-Guzik(2017)Gelbwaser-Klimovsky, and
  Aspuru-Guzik]{gelbwaser2017thermodynamic}
Gelbwaser-Klimovsky,~D.; Aspuru-Guzik,~A. \emph{Chemical Science}
  \textbf{2017}, \emph{8}, 1008--1014\relax
\mciteBstWouldAddEndPuncttrue
\mciteSetBstMidEndSepPunct{\mcitedefaultmidpunct}
{\mcitedefaultendpunct}{\mcitedefaultseppunct}\relax
\EndOfBibitem
\bibitem[Dubi and Di~Ventra(2009)Dubi, and Di~Ventra]{Dubi2009d}
Dubi,~Y.; Di~Ventra,~M. \emph{Nano Letters} \textbf{2009}, \emph{9},
  97--101\relax
\mciteBstWouldAddEndPuncttrue
\mciteSetBstMidEndSepPunct{\mcitedefaultmidpunct}
{\mcitedefaultendpunct}{\mcitedefaultseppunct}\relax
\EndOfBibitem
\bibitem[Manzano \latin{et~al.}(2012)Manzano, Tiersch, Asadian, and
  Briegel]{manzano2012}
Manzano,~D.; Tiersch,~M.; Asadian,~A.; Briegel,~H.~J. \emph{Physical Review E}
  \textbf{2012}, \emph{86}, 061118\relax
\mciteBstWouldAddEndPuncttrue
\mciteSetBstMidEndSepPunct{\mcitedefaultmidpunct}
{\mcitedefaultendpunct}{\mcitedefaultseppunct}\relax
\EndOfBibitem
\bibitem[Cho \latin{et~al.}(2005)Cho, Vaswani, Brixner, Stenger, and
  Fleming]{Cho2005}
Cho,~M.; Vaswani,~H.~M.; Brixner,~T.; Stenger,~J.; Fleming,~G.~R. \emph{The
  Journal of Physical Chemistry B} \textbf{2005}, \emph{109},
  10542--10556\relax
\mciteBstWouldAddEndPuncttrue
\mciteSetBstMidEndSepPunct{\mcitedefaultmidpunct}
{\mcitedefaultendpunct}{\mcitedefaultseppunct}\relax
\EndOfBibitem
\bibitem[Lloyd \latin{et~al.}(2011)Lloyd, Mohseni, Shabani, and
  Rabitz]{Lloyd2011}
Lloyd,~S.; Mohseni,~M.; Shabani,~A.; Rabitz,~H. \emph{arXiv preprint
  arXiv:1111.4982} \textbf{2011}, \relax
\mciteBstWouldAddEndPunctfalse
\mciteSetBstMidEndSepPunct{\mcitedefaultmidpunct}
{}{\mcitedefaultseppunct}\relax
\EndOfBibitem
\bibitem[Kassal and Aspuru-Guzik(2012)Kassal, and Aspuru-Guzik]{Kassal2012}
Kassal,~I.; Aspuru-Guzik,~A. \emph{New J. Phys.} \textbf{2012}, \emph{14},
  053041\relax
\mciteBstWouldAddEndPuncttrue
\mciteSetBstMidEndSepPunct{\mcitedefaultmidpunct}
{\mcitedefaultendpunct}{\mcitedefaultseppunct}\relax
\EndOfBibitem
\bibitem[Meixner(1965)]{Meixner1965}
Meixner,~J. Statistical mechanics of equilibrium and non-equilibrium.
  Statistical Mechanics of Equilibrium and Non-equilibrium. 1965\relax
\mciteBstWouldAddEndPuncttrue
\mciteSetBstMidEndSepPunct{\mcitedefaultmidpunct}
{\mcitedefaultendpunct}{\mcitedefaultseppunct}\relax
\EndOfBibitem
\bibitem[Kubo(1966)]{Kubo1966}
Kubo,~R. \emph{Reports on Progress in Physics} \textbf{1966}, \emph{29},
  255--284, cited By 2012\relax
\mciteBstWouldAddEndPuncttrue
\mciteSetBstMidEndSepPunct{\mcitedefaultmidpunct}
{\mcitedefaultendpunct}{\mcitedefaultseppunct}\relax
\EndOfBibitem
\bibitem[Gurvitz(2000)]{PhysRevLett.85.812}
Gurvitz,~S.~A. \emph{Phys. Rev. Lett.} \textbf{2000}, \emph{85}, 812--815\relax
\mciteBstWouldAddEndPuncttrue
\mciteSetBstMidEndSepPunct{\mcitedefaultmidpunct}
{\mcitedefaultendpunct}{\mcitedefaultseppunct}\relax
\EndOfBibitem
\bibitem[Diasi and Strunz(1997)Diasi, and Strunz]{diosi1997}
Diasi,~L.; Strunz,~W.~T. \emph{Physics Letters A} \textbf{1997}, \emph{235},
  569--573\relax
\mciteBstWouldAddEndPuncttrue
\mciteSetBstMidEndSepPunct{\mcitedefaultmidpunct}
{\mcitedefaultendpunct}{\mcitedefaultseppunct}\relax
\EndOfBibitem
\bibitem[Rebentrost \latin{et~al.}(2009)Rebentrost, Chakraborty, and
  Aspuru-Guzik]{rebentrost2009non}
Rebentrost,~P.; Chakraborty,~R.; Aspuru-Guzik,~A. \emph{The Journal of chemical
  physics} \textbf{2009}, \emph{131}, 11B605\relax
\mciteBstWouldAddEndPuncttrue
\mciteSetBstMidEndSepPunct{\mcitedefaultmidpunct}
{\mcitedefaultendpunct}{\mcitedefaultseppunct}\relax
\EndOfBibitem
\bibitem[Dutta and Bagchi(2017)Dutta, and Bagchi]{Dutta2017}
Dutta,~R.; Bagchi,~B. \emph{journal of physical chemistry letters}
  \textbf{2017}, \emph{8}, 5566\relax
\mciteBstWouldAddEndPuncttrue
\mciteSetBstMidEndSepPunct{\mcitedefaultmidpunct}
{\mcitedefaultendpunct}{\mcitedefaultseppunct}\relax
\EndOfBibitem
\bibitem[Eisenberg \latin{et~al.}(2014)Eisenberg, Yochelis, Ben-Harosh, David,
  Faust, Even-Dar, Taha, Haegel, Adir, Keren, and Paltiel]{Eisenberg2014}
Eisenberg,~I.; Yochelis,~S.; Ben-Harosh,~R.; David,~L.; Faust,~A.;
  Even-Dar,~N.; Taha,~H.; Haegel,~N.~M.; Adir,~N.; Keren,~N.; Paltiel,~Y.
  \emph{Phys. Chem. Chem. Phys.} \textbf{2014}, \emph{16}, 11196--11201\relax
\mciteBstWouldAddEndPuncttrue
\mciteSetBstMidEndSepPunct{\mcitedefaultmidpunct}
{\mcitedefaultendpunct}{\mcitedefaultseppunct}\relax
\EndOfBibitem
\bibitem[Eisenberg \latin{et~al.}(2017)Eisenberg, Harris, Levi-Kalisman,
  Yochelis, Shemesh, Ben-Nissan, Sharon, Raviv, Adir, Keren, \latin{et~al.}
  others]{eisenberg2017concentration}
others,, \latin{et~al.}  \emph{Photosynthesis Research} \textbf{2017},
  1--11\relax
\mciteBstWouldAddEndPuncttrue
\mciteSetBstMidEndSepPunct{\mcitedefaultmidpunct}
{\mcitedefaultendpunct}{\mcitedefaultseppunct}\relax
\EndOfBibitem
\bibitem[Banal \latin{et~al.}(2017)Banal, Kondo, Veneziano, Bathe, and
  Schlau-Cohen]{banal2017photophysics}
Banal,~J.~L.; Kondo,~T.; Veneziano,~R.; Bathe,~M.; Schlau-Cohen,~G.~S.
  \emph{The Journal of Physical Chemistry Letters} \textbf{2017}, \emph{8},
  5827--5833\relax
\mciteBstWouldAddEndPuncttrue
\mciteSetBstMidEndSepPunct{\mcitedefaultmidpunct}
{\mcitedefaultendpunct}{\mcitedefaultseppunct}\relax
\EndOfBibitem
\bibitem[Boulais \latin{et~al.}(2017)Boulais, Sawaya, Veneziano, Andreoni,
  Banal, Kondo, Mandal, Lin, Schlau-Cohen, Woodbury, \latin{et~al.}
  others]{boulais2017programmed}
others,, \latin{et~al.}  \emph{Nature Materials} \textbf{2017}, nmat5033\relax
\mciteBstWouldAddEndPuncttrue
\mciteSetBstMidEndSepPunct{\mcitedefaultmidpunct}
{\mcitedefaultendpunct}{\mcitedefaultseppunct}\relax
\EndOfBibitem
\bibitem[Penazzi \latin{et~al.}(2016)Penazzi, Pecchia, Gupta, and
  Frauenheim]{penazzi2016self}
Penazzi,~G.; Pecchia,~A.; Gupta,~V.; Frauenheim,~T. \emph{The Journal of
  Physical Chemistry C} \textbf{2016}, \emph{120}, 16383--16392\relax
\mciteBstWouldAddEndPuncttrue
\mciteSetBstMidEndSepPunct{\mcitedefaultmidpunct}
{\mcitedefaultendpunct}{\mcitedefaultseppunct}\relax
\EndOfBibitem
\bibitem[Kocherzhenko \latin{et~al.}(2010)Kocherzhenko, Grozema, and
  Siebbeles]{kocherzhenko2010charge}
Kocherzhenko,~A.~A.; Grozema,~F.~C.; Siebbeles,~L.~D. \emph{The Journal of
  Physical Chemistry C} \textbf{2010}, \emph{114}, 7973--7979\relax
\mciteBstWouldAddEndPuncttrue
\mciteSetBstMidEndSepPunct{\mcitedefaultmidpunct}
{\mcitedefaultendpunct}{\mcitedefaultseppunct}\relax
\EndOfBibitem
\bibitem[Nozaki \latin{et~al.}(2012)Nozaki, Da~Rocha, Pastawski, and
  Cuniberti]{nozaki2012disorder}
Nozaki,~D.; Da~Rocha,~C.~G.; Pastawski,~H.~M.; Cuniberti,~G. \emph{Physical
  Review B} \textbf{2012}, \emph{85}, 155327\relax
\mciteBstWouldAddEndPuncttrue
\mciteSetBstMidEndSepPunct{\mcitedefaultmidpunct}
{\mcitedefaultendpunct}{\mcitedefaultseppunct}\relax
\EndOfBibitem
\bibitem[Contreras-Pulido \latin{et~al.}(2014)Contreras-Pulido, Bruderer,
  Huelga, and Plenio]{contreras2014dephasing}
Contreras-Pulido,~L.; Bruderer,~M.; Huelga,~S.; Plenio,~M. \emph{New Journal of
  Physics} \textbf{2014}, \emph{16}, 113061\relax
\mciteBstWouldAddEndPuncttrue
\mciteSetBstMidEndSepPunct{\mcitedefaultmidpunct}
{\mcitedefaultendpunct}{\mcitedefaultseppunct}\relax
\EndOfBibitem
\bibitem[Xiang \latin{et~al.}(2015)Xiang, Palma, Bruot, Mujica, Ratner, and
  Tao]{xiang2015intermediate}
Xiang,~L.; Palma,~J.~L.; Bruot,~C.; Mujica,~V.; Ratner,~M.~A.; Tao,~N.
  \emph{Nature chemistry} \textbf{2015}, \emph{7}, 221--226\relax
\mciteBstWouldAddEndPuncttrue
\mciteSetBstMidEndSepPunct{\mcitedefaultmidpunct}
{\mcitedefaultendpunct}{\mcitedefaultseppunct}\relax
\EndOfBibitem
\end{mcitethebibliography}
%\end{document}
\providecommand{\latin}[1]{#1}
\makeatletter
\providecommand{\doi}
  {\begingroup\let\do\@makeother\dospecials
  \catcode`\{=1 \catcode`\}=2\doi@aux}
\providecommand{\doi@aux}[1]{\endgroup\texttt{#1}}
\makeatother
\providecommand*\mcitethebibliography{\thebibliography}
\csname @ifundefined\endcsname{endmcitethebibliography}
  {\let\endmcitethebibliography\endthebibliography}{}

\end{document}